%% file: main.tex
\documentclass[sigconf]{acmart}

\usepackage{amsmath,amsfonts}
\usepackage{graphicx}
\usepackage{xcolor}
\usepackage{comment}
\usepackage{tikz}
\usetikzlibrary{shapes,arrows.meta,positioning}

\usepackage[textsize=footnotesize, textwidth=15mm]{todonotes}
\presetkeys{todonotes}{fancyline}{}
\usepackage[inline]{enumitem}
\usepackage[english=american,babel=false]{csquotes}
\usepackage{url}
\usepackage{hyperref}

\setcopyright{none}
\acmConference[LIMITS '26]{12th Workshop on Computing within Limits}{June 23--25, 20}{Online}
\acmDOI{10.48550/arXiv.2606.06181}
\acmISBN{}

\title{Opportunities and Challenges in Securely Reusing and Repurposing Mobile Devices}

\author{Adelin Roty}
\affiliation{
  \institution{\'Ecole Polytechnique de Bruxelles,
    BEAMS, Universit\'e Libre de Bruxelles}
  \city{Brussels}
  \country{Belgium}
}
\email{adelin.roty@ulb.be}

\author{Jan Tobias M\"uhlberg}
\affiliation{
  \institution{\'Ecole Polytechnique de Bruxelles,
    BEAMS, Universit\'e Libre de Bruxelles}
  \city{Brussels}
  \country{Belgium}
}
\email{jan.tobias.muehlberg@ulb.be}

\author{Jean-Fran\c{c}ois Determe}
\affiliation{
  \institution{\'Ecole Polytechnique de Bruxelles,
    BEAMS, Universit\'e Libre de Bruxelles}
  \city{Brussels}
  \country{Belgium}
}
\email{jean-francois.determe@ulb.be}

\begin{abstract}
\input{abstract.tex}

\end{abstract}

\input{ccs.tex}


\begin{document}

\maketitle

\input{intro.tex}

\input{background.tex}

\input{pinephoneCS.tex}

\input{experiments.tex}

\input{discussion.tex}

\input{conclusions.tex}

\balance
\input{acks.tex}

\bibliographystyle{ACM-Reference-Format}
\bibliography{zotero-thesis_aroty.bib,non-zotero.bib}

\end{document}

%% file: abstract.tex
An estimated 5.3 billion mobile phones became electronic waste in 2022.
Many of these devices can be repurposed and used in different contexts to extend their lifetime and to reduce ecological impacts.
An often overlooked aspect of smartphone reuse is cybersecurity: these devices embed hardware-backed security mechanisms that rely on vendor-controlled provisioning and are designed for a fixed device lifecycle.
In this paper, we investigate whether security mechanisms and guarantees remain effective when devices are repurposed outside their original ecosystem.
We explore security features in a PinePhone, an open-hardware smartphone, and focus on three core security aspects: boot chain integrity, isolation provided by the Trusted Execution Environment, and the protection of hardware-bound secrets.
Our experiments simulate realistic repurposing scenarios and highlight the complexity of reconstructing trust anchors.
We generalize our observations to infer requirements for secure repurposing and illustrate how vendor-locked mechanisms hinder the repurposing of a majority of discarded devices.

%% file: ccs.tex
\ccsdesc[500]{Security and privacy~Software and application security}
\ccsdesc[500]{Security and privacy~Mobile platform security}
\ccsdesc[500]{Security and privacy~Trusted computing}
\ccsdesc[500]{Security and privacy~Embedded systems security}

%% file: intro.tex
\section{Introduction}
\label{sec:intro}

The rapid turnover of smartphones contributes significantly to the growing
volume of electronic waste. Estimates by the WEEE Forum indicate that
5.3 billion mobile phones became waste in
2022~\cite{charytanowicz_international_2022}.
While recycling of electronic waste focuses on extracting raw materials,
many discarded devices actually remain functional, either at the level of
individual components such as processors or cameras, or as a whole.
Extending the lifetime of these devices through repurposing has therefore
emerged as a promising approach to reduce environmental impact and improve
resource efficiency.

Beyond their original use, smartphones are increasingly reused for a
variety of applications, ranging from low-cost computing nodes
and sensing platforms to communication gateways or display units. Their
high level of integration, combining powerful processors, wireless
connectivity, and diverse sensors, makes them attractive for
repurposing, particularly in embedded and Internet-of-Things
(IoT) contexts. The \enquote{Computing within Limits} community has
engaged with repurposing and upcycling electronics in its own
creative ways. More recently, Lafr\'echoux discusses pathways towards
retrofitting devices and explores the resulting hybrid interaction
patterns, such as monotasking on dedicated hardware with intermittent
connectivity~\cite{lafrechoux_practical_2025}. Rigaud explores the
possibilities of reusing obsolete smartphones and tablets to build
interactive musical instruments and discusses opportunities towards
developing an open-source toolkit that would enable anyone to build new
interactive systems from old devices~\cite{rigaud_zombitron_2025}.
Furthermore, Stojanov et al.~\cite{stojanov_how_2023} and Snodgrass et
al.~\cite{snodgrass_windternet_2024} engage with requirements and design
practices for off-grid computing from a permacomputing perspective, based
on local energy generation, the use of lightweight control elements, and
Raspberry-Pi-like equipment for servers, with an emphasis on reused
electronics. Repurposing smartphones fits well within this context, as
modern mobile devices are designed for low power consumption and typically
provide computing resources beyond those of a Raspberry Pi. They can thus
be used to monitor and control energy generation and battery charging, or
to host reasonably lightweight compute and internet services.

Outside of the LIMITS community, the reuse potential of smartphones is
also well recognized: Switzer et al.~\cite{switzer_junkyard_2023}
introduce the term \enquote{junkyard computing} and conclude that for
\enquote{specific workloads, clusters of repurposed phones are cheaper and
more carbon efficient than traditional servers.} Research also shows that
repair, refurbishing, and reuse of smartphones shows the highest potential
for impact reduction across several
categories~\cite{bieser_lifetime_2021,pamminger_modelling_2021}. This
potential may be limited by reliability and cost-effective repairability
options for older devices, and manufacturers should implement design choices
that foster longevity of devices~\cite{cordella_durability_2021}. Overall,
we see a range of initiatives across different communities that aim to
develop low-impact computing facilities, \enquote{elevating sustainability to a first-class consideration for computing and computing facility
design and implementation decisions}~\cite{knowles_climate_2025}.

One important aspect in smartphone reuse remains largely overlooked in the
literature: cybersecurity and the links between cybersecurity and privacy
and safety of the reused device in a new operational context. Modern
smartphones embed hardware-backed security mechanisms, including secure
boot and Trusted Execution Environments (TEEs), which are designed to
enforce system integrity, isolate sensitive operations, and protect
cryptographic material and user data. These mechanisms are typically
deployed within tightly controlled provisioning models and assume a
predictable device lifecycle and a trusted software ecosystem.
Arm processors are the most widely used CPUs in the mobile and embedded
markets, and come with Arm TrustZone TEE
support~\cite{pinto_demystifying_2019}. TrustZone separates the device into
two worlds: a non-secure Rich Execution Environment (REE) where, e.g., the
Android operating system runs; and a secure TEE where a vendor-specific
operating system is executing. Security-critical applications such as the
Android Keystore~\cite{google_hardware-backed_2026} use hardware-backed
cryptography through a Hardware Abstraction Layer (HAL) that is implemented
by the vendor-specific TEE. In many smartphones, using TEE functions
requires permission from the phone vendor, replacing the TEE layer is close
to impossible, and, of course, TEEs can have vulnerabilities or be
compromised, and updates and fixes also depend on the phone
vendor~\cite{pinto_demystifying_2019,shakevsky_trust_2022}.

However, repurposing a device often requires replacing core software
components, such as the operating system or the bootloader, to increase
flexibility and control. Inevitably, such actions disrupt the original
security architecture while, from most vendors' perspective and business
model, there is little to no concern or support to conserve hardware-backed
security guarantees when a device is removed from its initial ecosystem.

\enlargethispage{1em}
\paragraph{This Paper \& Contributions}
In this paper, we study hardware-backed security in repurposed smartphones.
We seek to understand how security features can be turned into an asset the
enables trust in critical reuse cases, or if they pose a liability because
they remain vendor-locked, cannot be used, or enable third parties to
interfere with repurposing. Since security features are implemented
differently across devices, we focus on the Pine64 PinePhone, an
open-hardware platform based on an Arm Cortex-A7 processor with TrustZone
support, integrated in a common System-on-Chip (SoC). We develop a case
study that explores the limits of secure repurposing under favorable
conditions, from which we generalize to more vendor-controlled scenarios.

To guide our investigation, we define four core research objectives: (i)
evaluating the integrity of the boot chain, (ii) assessing the isolation
guarantees of the Trusted Execution Environment (TEE), (iii) examining
the protection of hardware-bound secrets, and (iv) understanding the
applicability of repurposed smartphones in security-critical scenarios.
Through controlled experiments, we simulate realistic repurposing
conditions and analyze to what extent these properties can be preserved,
degraded, or reconstructed.

Our results show that achieving meaningful hardware-backed security in
repurposed devices requires reconstructing trust anchors at a low level.
Even on an open platform, this process is complex and error-prone,
highlighting structural challenges for secure smartphone reuse. These
findings expose a fundamental gap between hardware capabilities and
enforceable security guarantees in repurposed devices, suggesting that
secure reuse cannot rely solely on existing platform mechanisms but
requires explicit re-establishment of trust anchors. Our main conclusion is
that, to meaningfully engage with ideas for repurposing the vast quantity
of discarded mobile phones, active support by device vendors is required to
turn hardware-backed security into an asset that can support diverse reuse
scenarios. This likely requires stricter and more comprehensive
regulations, e.g., in the context of an extended Right to Repair, to be put
in place.

\paragraph{Organization}
The remainder of this paper is organized as follows: Section~\ref{sec:background} presents
background and motivation.  Section~\ref{sec:pinephone} details our case study based on a
PinePhone, while Section~\ref{sec:experiments} details our experiments and findings, followed
by a discussion in Section~\ref{sec:discussion}. We summarize our work and draw final
conclusions in Section~\ref{sec:conclusions}. Supplementary materials,
including source code is made available as an accompanying artifact to this
paper.

%% file: background.tex
\section{Background and Motivation}
\label{sec:background}

\subsection{The Environmental Case for Full Device Reuse}

Each year, an estimated five billion smartphones are discarded worldwide~\cite{bbc_e-waste_2022,charytanowicz_international_2022}, despite many remaining fully functional~\cite{martinho_consumer_2017,cordella_reducing_2021}.
The average lifetime of a smartphone is around three years~\cite{noauthor_revealed_2019}, after which devices are often replaced even when their hardware capabilities remain sufficient.
Consumer behavior studies report that a significant fraction of these devices is not immediately recycled, but instead stored unused in households or informal environments~\cite{martinho_consumer_2017}.

Among the devices that do enter formal recycling processes~\cite{forti_global_nodate,balde_global_2024}, material recovery provides only partial mitigation of environmental impact.
Although recycling technologies continue to improve, material recovery alone does not compensate for the environmental cost of manufacturing complex electronic devices~\cite{cenci_composition_2023}: the recovery of high-quality materials remains limited~\cite{gulliani_recovery_2023,van_yken_e-waste_2021}, while requiring substantial energy and industrial processing~\cite{hu_carbon_2017,he_life_2020}.

In contrast, circular economy strategies such as repair,
refurbishment, and remanufacturing aim to extend device lifetimes
and preserve functional components~\cite{bigliardi_environmentally-conscious_2022}.
Life-cycle assessment studies indicate that maintaining devices in
operational use phases significantly reduces their overall
environmental footprint compared to immediate recycling or
disposal~\cite{pamminger_modelling_2021}.
This suggests that full-device reuse, rather than dismantling,
offers a more effective approach to reducing environmental impact.
However, the implications of such extended lifetimes for
hardware-backed security remain largely under-explored,
particularly when devices are repurposed outside their original
ecosystem. In particular, little attention has been paid to
whether the hardware-backed security properties of smartphones
can be preserved in such repurposing scenarios.

\subsection{Smartphones as High-Capability Embedded Platforms}

Modern smartphones integrate highly capable system-on-chips
(SoCs), typically featuring multi-core processors clocked in the
GHz range, several gigabytes of RAM, and high-throughput storage
interfaces.
Recent work on \enquote{junkyard computing} explicitly highlights
the untapped potential of discarded smartphones as a large-scale
computational resource, demonstrating that even older devices can
support modern workloads~\cite{switzer_junkyard_2023}.
They also embed multiple wireless communication interfaces,
including cellular, Wi-Fi, and Bluetooth, as well as a wide range
of sensors and dedicated hardware cryptographic accelerators.

This level of integration and performance places smartphones on
par with, or even above, many embedded and IoT platforms,
motivating interest in reusing or upcycling discarded
devices~\cite{reshwanth_iot_2023,iera_making_2011,ward_building_2018}.

Beyond general-purpose computation, commodity smartphones
incorporate hardware-backed security mechanisms such as secure
boot and Trusted Execution Environments (TEEs)~\cite{schneider_sok_2022},
specifically the TrustZone~\cite{pinto_demystifying_2019} and
CCA~\cite{arm_arm_2021,huang_sok_2024} features of Arm processors,
which are commonly used in mobile devices.
These TEE mechanisms are designed to enforce system integrity,
isolate sensitive operations, and protect cryptographic key
material.
Hardware-backed TEEs follow common architectural principles to
enable verifiable launch, run-time isolation, and secure
storage~\cite{pinto_demystifying_2019,schneider_sok_2022}.
Research has shown that TrustZone-based systems can provide
real-time and availability guarantees~\cite{wang_rt-tee_2022} and
can be used to balance the control different stakeholders exert
over devices, e.g., enabling notions of sovereignty~\cite{groschupp_its_2023}
in mobile phones, and further demonstrate how
hardware-enforced isolation primitives can support secure system
design.

While these works do not specifically target smartphone
repurposing, they highlight that commodity processors—often
architecturally similar to those used in mobile platforms—are
capable of supporting strong hardware-enforced security
guarantees~\cite{schneider_sok_2022,wang_rt-tee_2022,groschupp_its_2023}.
This suggests that, in principle, repurposed smartphones could
serve as secure embedded platforms, provided that their hardware
security features remain accessible and correctly configured
after system modification.

\subsection{Hardware-Backed Security and Platform Control}
\label{subsec:hwsec_control}

In mainstream smartphone deployments, hardware-backed security mechanisms are tightly coupled with vendor-controlled provisioning models.
Secure boot chains are typically anchored in immutable hardware roots of trust, such as on-chip ROM code and fused cryptographic keys, provisioned during manufacturing.
These mechanisms enforce a verified boot sequence from early firmware stages to the operating system, preventing the execution of unauthorized code.
In practice, a large majority of commercial smartphones implement some form of verified boot and secure key storage.

TEEs, commonly implemented using ARM TrustZone technology, provide isolated execution contexts for sensitive services such as authentication, key management, and attestation~\cite{schneider_sok_2022}.
These environments rely on hardware-enforced separation between secure and non-secure worlds, and are typically integrated with vendor-specific software stacks and provisioning infrastructure.

However, authority over these security anchors remains largely under the control of device manufacturers and operating system vendors.
Cryptographic keys, boot policies, and TEE configurations are usually provisioned at production time and cannot be modified without breaking the original trust model.
Proposals advocating for \enquote{sovereign} smartphone architectures highlight this structural asymmetry, arguing that users lack meaningful control over the hardware root of trust and associated provisioning mechanisms~\cite{groschupp_sovereign_2021}.
Even when bootloader unlocking is supported, it often disables or bypasses critical security features such as verified boot or key attestation.

This creates a fundamental tension for repurposing:
extending the lifetime of a device typically requires modifying or replacing system software components.
Yet, hardware-backed security mechanisms are explicitly designed to prevent such modifications unless they are authorized within the original provisioning model.
As a result, increasing software freedom often comes at the cost of weakened or entirely disabled hardware-enforced security guarantees.

\subsection{Research Gap}

Existing literature extensively addresses circular economy strategies for smartphones~\cite{pamminger_modelling_2021} and the architectural design of hardware-supported TEEs~\cite{schneider_sok_2022,groschupp_its_2023}.

Prior work on smartphone durability highlights that limited software support and update lifecycles provided by manufacturers are a key factor limiting device reuse, as devices rapidly become outdated or insecure once official support ends~\cite{cordella_durability_2021}. This has motivated both regulatory discussions and industry efforts toward extending software maintenance periods, with some manufacturers now providing longer update guarantees for selected device lines.

Beyond industrial approaches, a broader ecosystem of right-to-repair initiatives and open-source mobile platforms has emerged to address smartphone obsolescence from a different angle. Projects such as community-driven operating systems (Ubuntu Touch~\cite{ubuntu_touch}, Mobian~\cite{mobian}, postmarketOS~\cite{postmarketos},...) and advocacy groups such as the Software Freedom Conservancy~\cite{SFC} promote user autonomy, repairability, and long-term software support for consumer devices. Similarly, emerging initiatives in the open hardware and mobile sovereignty space like LibrePhone~\cite{librephone} aim to reduce dependence on vendor-controlled software lifecycles by enabling alternative software stacks and user-controlled deployments.
These efforts demonstrate that practical pathways exist for extending the functional lifetime of smartphones beyond manufacturer-defined end-of-life, often by prioritizing repair, reuse, and software freedom.

However, these initiatives primarily focus on software longevity, repairability, and user control over device functionality. They generally do not address whether hardware-backed security mechanisms remain usable, trustworthy, or correctly configured once devices are repurposed outside their original provisioning and certification ecosystems.

Conceptual proposals further explore how control over smartphone security architectures could be re-balanced toward users~\cite{groschupp_sovereign_2021}. These works argue for greater user sovereignty over hardware roots of trust and provisioning mechanisms, but remain largely at a design or conceptual level rather than evaluating real-world feasibility on repurposed devices.

To the best of our knowledge, there is no empirical work evaluating whether hardware-backed security guarantees remain usable in practice once a commodity smartphone is repurposed outside its original provisioning and update ecosystem. In particular, prior work does not systematically assess how secure boot mechanisms behave after modification, whether TEEs remain operational under alternative software stacks, or how hardware roots of trust can be accessed or re-established in such scenarios.

This gap motivates a structured empirical case study of secure smartphone repurposing on an open platform, with the objective of identifying both practical capabilities and structural limitations.

%% file: pinephoneCS.tex
\section{Case Study on a Pinephone}
\label{sec:pinephone}

In this section, we present our case study on the PinePhone, a
TrustZone-enabled open hardware smartphone. We first provide an
overview of the platform and its security architecture, highlighting
features relevant to hardware-backed security. We then describe the
repurposing context considered in our experiments, including threat
assumptions and security-relevant characteristics that influence the
enforcement of hardware mechanisms. Finally, we discuss the
limitations of our study, clarifying which findings can be generalized
and which are specific to this platform.
\subsection{Platform Overview}

The PinePhone is an open hardware smartphone platform based on the Allwinner A64 system-on-chip (SoC). Unlike mainstream commercial devices, it emphasizes transparency, modifiability, and developer accessibility~\cite{noauthor_pinephone_nodate,noauthor_pinephone_nodate-1}.
These characteristics make it a suitable platform for controlled security experimentation, while still representing a realistic Arm-based mobile architecture. In addition, the Allwinner ecosystem benefits from an active open-source community, providing documentation and software support~\cite{noauthor_pine64_nodate}.

The Allwinner A64 integrates a quad-core Arm Cortex-A53 processor implementing the ARMv8-A architecture with TrustZone extensions. 
It supports both secure and non-secure execution environments, enabling hardware-enforced privilege separation.

The device typically boots either from eMMC or a microSD card.
This flexible boot configuration is particularly relevant in repurposing scenarios, as system images can be easily modified, replaced, or redirected to alternative storage media.

\subsection{Security Architecture}
\label{subsec:secarch}

The boot process of the Allwinner A64 begins in an immutable BootROM embedded in the SoC.
This stage performs minimal hardware initialization and loads a first-stage bootloader program from a predefined storage medium.
This program is usually a derivative of the open-source project U-Boot Secondary Program Loader (SPL) \cite{noauthor_booting_nodate}.

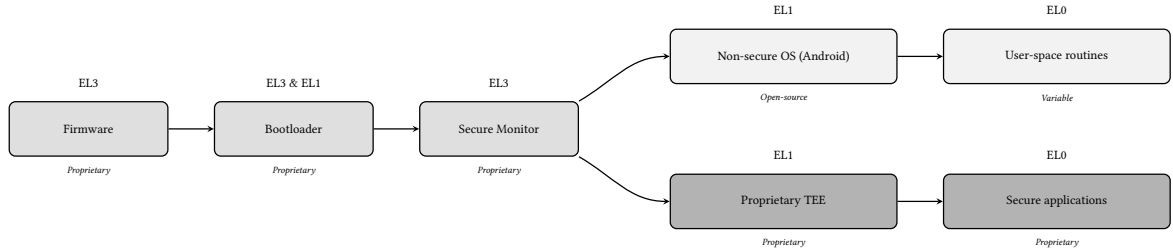
\begin{figure*}[tb]
    \centering
    \scalebox{0.60}{\input{bc_init}}
    \caption{Typical Smartphone Boot-chain}
    \label{fig:bc_init}
\end{figure*}

The SPL is responsible for complex hardware initialization (including DRAM) and subsequently loads the full U-Boot bootloader (U-Boot proper), which in turn loads the Linux kernel and associated device tree from persistent storage.
This establishes a classical multi-stage boot chain from immutable ROM to the operating system (see figure~\ref{fig:bc_init}).

The BootROM does not support cryptographic verification of the boot chain, but the SPL itself can be configured to check the following steps of the boot chain.
On the PinePhone, a form of secure boot \cite{noauthor_secureboot_nodate,noauthor_verified_nodate} is not enforced by default, and the integrity of the boot chain depends on software configuration choices rather than immutable provisioning.

The A64 also implements Arm TrustZone, which partitions system resources into secure and non-secure domains.
At the hardware level, this includes CPU execution modes (EL3), memory isolation through the TrustZone Address Space Controller (TZASC), and differentiated peripheral access control~\cite{noauthor_trustzone_nodate,pinto_demystifying_2019}.

Secure world execution is typically mediated by a Secure Monitor (SM), which handles transitions between execution domains using SM Calls~\cite{noauthor_arm_nodate,noauthor_tf-overview_nodate}.
A TEE is deployed on top of this infrastructure to provide isolated services for sensitive operations.

\subsection{Repurposing Context}

In our study, the PinePhone is considered in a repurposed
scenario where the original operating system and firmware
are replaced, and the device is reconfigured for a new
operational context.
This reflects realistic reuse patterns in open hardware
and sustainability-driven environments, where devices are
removed from their original vendor-controlled ecosystem.

As stated in section \ref{subsec:hwsec_control}, repurposing typically involves unlocking the bootloader to allow custom firmware modification, and deploying a custom software stack.
This transition fundamentally alters the security model: once the bootloader is unlocked, hardware-backed verification mechanisms are often disabled or bypassed, and the original chain of trust is no longer enforced.
Even though the PinePhone does not require bootloader unlocking, the replacement of the stock bootloader is sufficient to disable the original platform hardware security capacities.

\subsubsection{Threat Model}

We consider a powerful software adversary with control over
the non-secure operating system environment.
This includes the ability to execute arbitrary code with
high privileges and to modify persistent storage.

We do not assume invasive hardware attacks, such as physical decapsulation or fault injection.
Our focus is on evaluating whether hardware-backed security mechanisms remain enforceable during a repurposing scenario.

\subsubsection{Security-Relevant Characteristics}

Repurposing introduces specific conditions that directly
impact the enforcement of hardware-backed security guarantees.

\textbf{Loss of enforced boot integrity.}
Unlocking the bootloader typically disables secure boot mechanisms.
As a result, the integrity of the boot chain is no longer checked by default.
Unless secure boot is explicitly configured and private keys are provisioned into a specific SoC memory area, arbitrary boot components can be loaded from writable storage. 

\textbf{TrustZone availability without policy binding.}
While the hardware continues to support secure and non-secure execution, the presence and configuration of a Trusted Execution Environment depend entirely on the deployed software stack.
Isolation guarantees are therefore not inherent to the platform, but must be explicitly implemented.

These characteristics illustrate a fundamental shift: hardware
security features remain available, but their effective
enforcement becomes contingent on the correctness and
completeness of the repurposed software stack.
It is up to the system integrator to leverage those mechanisms again.

\subsection{Limitations of our study}

While the PinePhone provides valuable insight into TrustZone-capable hardware reuse, it differs from mainstream commercial smartphones in several ways.

First, the manufacturer keys used during secure boot checks are not fused in memory during production, allowing the end-user to replace them with custom ones if needed.
Second, it does not enforce mandatory verified boot with rollback protection in its default configuration.
Third, it easily allows for bootloader unlocking, which is not a characteristic shared by all mainstream smartphones

In addition, the PinePhone is a comparatively open platform, with extensive publicly available documentation and active community support.
This significantly lowers the barrier for firmware modification and system repurposing, in contrast to mainstream devices where such documentation is often limited or unavailable.

As a result, findings obtained on the PinePhone may not directly generalize to more tightly controlled, vendor-locked smartphones, where access to low-level components and security mechanisms is more restricted.

However, these characteristics enable controlled experimentation at the boundary between configurable and hardware-enforced security properties. In particular, they allow us to isolate the role of software configuration in re-establishing security guarantees after repurposing.

Overall, the PinePhone represents a TrustZone enabled Arm platform where secure boot and isolation mechanisms are technically feasible but not rigidly enforced.
This makes it a suitable case study for evaluating which hardware-backed security properties survive repurposing and which depend critically on original provisioning assumptions.
As a result, this work---while interesting on its own---can also be portrayed as a stepping stone towards repurposing less open smartphones.

%% file: bc_init.tex
\begin{tikzpicture}[
  font=\small,
    inibox/.style={
    draw,
    thick,
    rounded corners,
    minimum width=3.5cm,
    minimum height=1.2cm,
    align=center,
    fill=gray!25
  },
  rbox/.style={
    draw,
    thick,
    rounded corners,
    minimum width=5cm,
    minimum height=1.2cm,
    align=center,
    fill=gray!10
  },
  secbox/.style={
    draw,
    thick,
    rounded corners,
    minimum width=5cm,
    minimum height=1.2cm,
    align=center,
    fill=gray!60
  },
  labelstyle/.style={
    font=\scriptsize\itshape,
    align=center
  },
  >=stealth
]

% Étapes horizontales
\node[inibox] (fw) {Firmware};
\node[inibox, right=1cm of fw] (bl) {Bootloader};
\node[inibox, right=1cm of bl] (sm) {Secure Monitor};

% Split en deux colonnes après Secure Monitor, décalées vers la droite
\node[rbox, right=2cm of sm, yshift=1.6cm] (os) {Non-secure OS (Android)};
\node[secbox, right=2cm of sm, yshift=-1.6cm] (tee) {Proprietary TEE};

% User-space / Secure applications
\node[rbox, right=1cm of os] (usr) {User-space routines};
\node[secbox, right=1cm of tee,] (secapp) {Secure applications};

% Flèches avec courbes plus naturelles
\draw[->, thick] (fw.east) -- (bl.west);
\draw[->, thick] (bl.east) -- (sm.west);
\draw[->, thick] (sm.north east) to[out=30,in=180] (os.west);
\draw[->, thick] (sm.south east) to[out=-30,in=180] (tee.west);
\draw[->, thick] (os.east) -- (usr.west);
\draw[->, thick] (tee.east) -- (secapp.west);

\node[label, above=2mm of fw.north] {EL3};
\node[label, above=2mm of bl.north] {EL3 \& EL1};
\node[label, above=2mm of sm.north] {EL3};
\node[label, above=2mm of os.north] {EL1};
\node[label, above=2mm of tee.north] {EL1};
\node[label, above=2mm of usr.north] {EL0};
\node[label, above=2mm of secapp.north] {EL0};

\node[labelstyle, below=1mm of fw.south] {Proprietary};
\node[labelstyle, below=1mm of bl.south] {Proprietary};
\node[labelstyle, below=1mm of sm.south] {Proprietary};
\node[labelstyle, below=1mm of os.south] {Open-source};
\node[labelstyle, below=1mm of tee.south] {Proprietary};
\node[labelstyle, below=1mm of usr.south] {Variable};
\node[labelstyle, below=1mm of secapp.south] {Proprietary};

\end{tikzpicture}

%% file: experiments.tex
\section{Experiments and Findings}
\label{sec:experiments}

In this section, we present the sequence of experiments conducted on the PinePhone to evaluate the feasibility of repurposing a device while preserving hardware-backed security properties.
The focus is on the boot chain, the potential integration of a TEE, and the implications for hardware-bound secrets. Each stage highlights which endeavors were successfully achieved (and how), encountered challenges, and the resulting status of the boot chain.

\subsection{Target Software Stack}

As stated in section \ref{subsec:secarch}, the PinePhone does not implement a locked bootloader model.
Instead, the system boots directly from external storage, leaving secure booting capacities in the sole hands of the bootloader.

We identified those different components to be mandatory to successfully retrieve the security characteristics we are targeting in our repurposed smartphone:
\begin{itemize}
	\item{A bootloader with secure booting capacities}
	\item{A SM} to manage context switching between TrustZone secure and non-secure world
	\item{A Linux kernel} to manage the non-secure userland
	\item{A TEE} to simplify usage of the hardware security primitives in the secure world
\end{itemize}

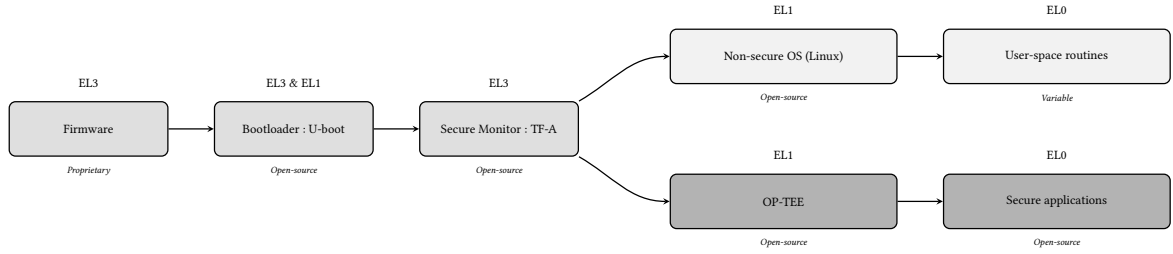
\begin{figure*}[tb]
	\centering
	\scalebox{0.60}{\input{bc_typ}}
	\caption{Targeted Software Stack for the PinePhone in our experiments}
	\label{fig:bc_typ}
\end{figure*}

The integration of those different elements together, and where they start to exist in the boot process are presented in figure \ref{fig:bc_typ}.

It is important to note that both the Firmware and U-boot transfer full control to the next element.
On the other hand the SM, kernel and TEE coexist and interact during the rest of the device usage.

\subsection{Flashing a Minimal U-Boot Image}
\label{subsec:ubootimg}

The first experiment consisted in flashing an image containing only the U-Boot program compiled from source (including both the SPL and U-Boot proper).
This experiment aimed at validating the earliest controllable stage of the boot chain in a repurposed context, and set up a working framework for more complex experiments.

\begin{figure}[tb]
    \centering
    \includegraphics[width=0.5\linewidth]{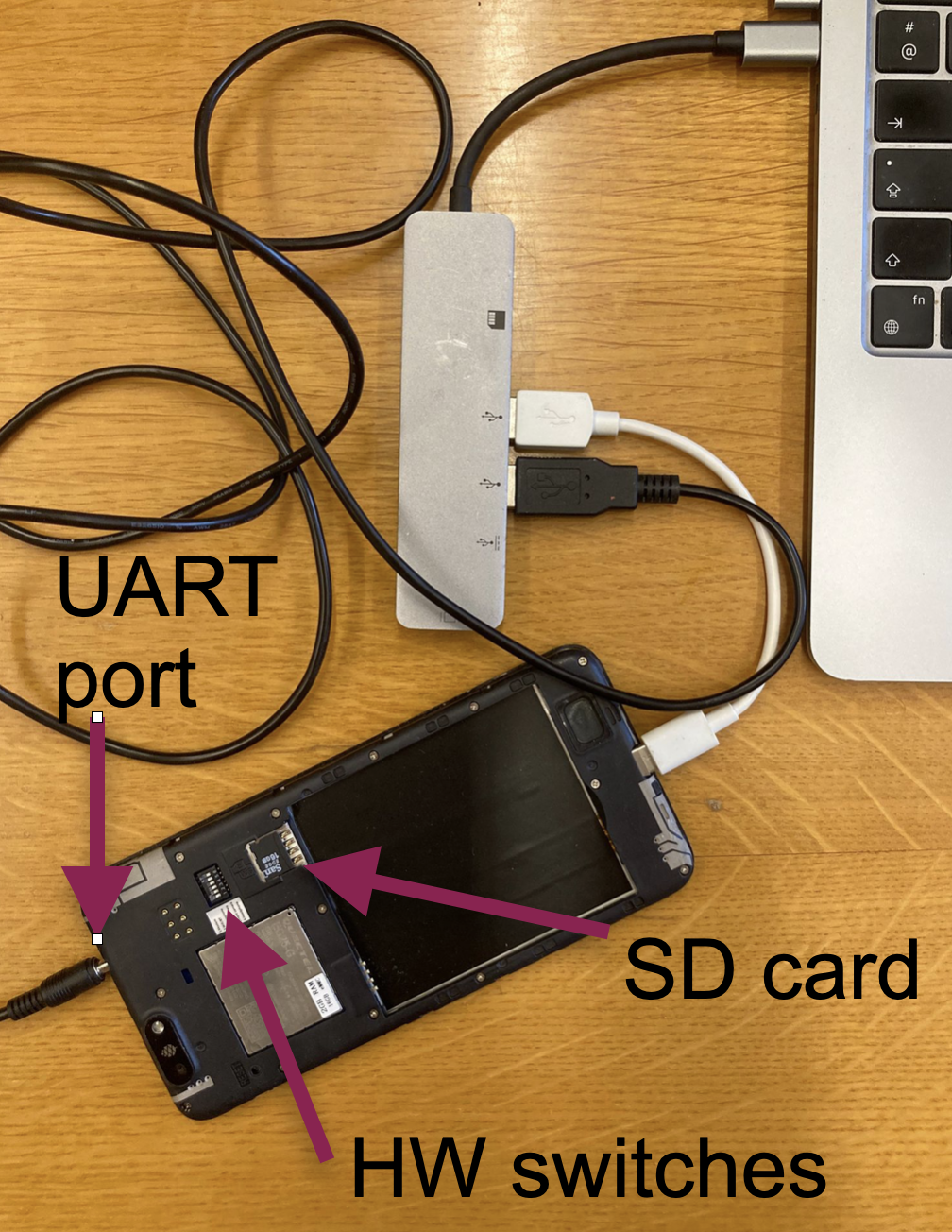}
    \caption{Setup to Interact with a Headless PinePhone}
    \label{fig:expsetup}
\end{figure}

Through hardware switches at the back of the phone (see figure \ref{fig:expsetup}, the PinePhone enables its audio jack to be used as an UART interface~\cite{pine_privacy_hw}.
This port will reveal to be of great importance, because it conveniently allows to retrieve startup logs on a host computer, making it easier to backtrack possible boot errors and check the system state.

The creation of the U-Boot image required on the host computer required the usage of a cross-compiler (see\cite{crosscompuboot}) and parameterization of compile options.
One of the critical parameters to set up is the Device Tree Source (DTS): an object file thoroughly describing the hardware of the targeted platform.
Thanks to the openness of the PinePhone, its DTS is accessible in the Linux Kernel source code, but this is not the case for all smartphones.
This image was afterward directly written on the SD card used by the PinePhone to boot.

\begin{figure}[tb]
    \centering
    \includegraphics[width=0.7\linewidth]{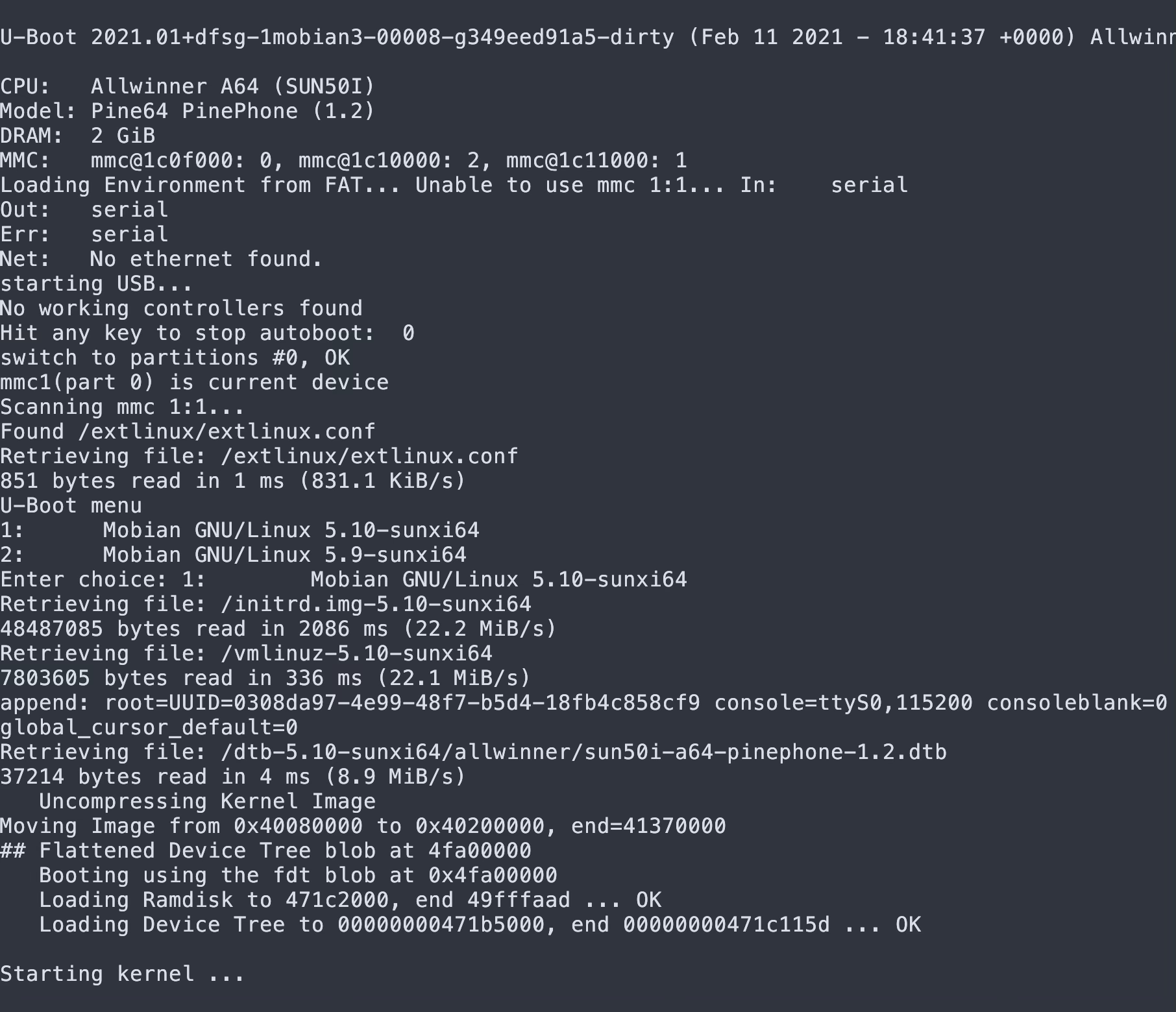}
    \caption{Logs of a Successful U-Boot launch}
    \label{fig:ubootlog}
\end{figure}

An example of the collected boot logs is presented in figure \ref{fig:ubootlog}, showing that U-Boot started up properly and was trying to continue the boot process, i.e. launch a kernel not existent in the current image.

While U-Boot supports optional secure boot mechanisms through signature verification of subsequent stages, this feature was not enabled at this stage of the experiments, as later components of the boot chain were not yet in place.
More importantly, even if such mechanisms were configured, they would not provide a complete secure boot guarantee in this context.
Since the BootROM does not verify the authenticity of the SPL, an attacker can replace the entire bootloader with a modified version, thereby bypassing any verification implemented at later stages.

\subsection{Secure Monitor and Kernel Launch}

The direct extension of the successful image and workflow presented in section \ref{subsec:ubootimg} is the inclusion of a SM and a Linux kernel.

Compilation of the SM and kernel did not require additional information (having the DTS is mandatory, but it was already the case for U-Boot), and the same cross-compilation toolchain could be used.

Upon powering the device, the BootROM again successfully loaded U-Boot. Unlike the previous experiment, it was now able to locate a valid Linux image on the storage medium and proceeded to load and execute the kernel.

The boot process completed successfully, resulting in a functional Linux environment running on the PinePhone.
The system was sufficiently stable to support interactive use and basic regular operations such as SSH connection through the serial port, interaction with the filesystem, basic scripting,...

This setup is representative of regular smartphones recycling projects, where the device original software stack is replaced with a custom bootloader and a Linux image to provide a generic embedded system.
Once booted, standard package managers and tools can be used to adapt the system to a wide variety of applications, ranging from IoT gateways to low-cost servers or sensor platforms.
Compared to figure \ref{fig:bc_typ}, only the \emph{top part} (non-secure world) of the software stack is implemented, which does not enable the leverage of the SoC TrustZone capacities.
As a result, sensitive operations and cryptographic material cannot be protected against a compromised operating system.

Furthermore, although U-Boot supports optional verification mechanisms like secure boot, these were not enabled in this configuration.
The lack of a hardware-enforced root of trust allows an attacker to replace any stage of the boot chain, including the bootloader itself, thereby bypassing any software-based integrity checks.
While software-based protections such as filesystem or disk encryption can be implemented at the operating system level to protect user data, they do not provide equivalent guarantees.

\subsection{Image Generation through Buildroot}

To move toward a more reproducible and extensible workflow, Buildroot was used to generate the same complete image for the PinePhone.
This experiment aimed to provide a generic framework for deploying similar systems across multiple platforms.

Buildroot is a tool for automatically building a complete Linux system for embedded targets \cite{buildroot}.
It generates a cross-compilation tool chain, bootloader, kernel and root filesystem and allows for easier integration of additional packages of functionalities, making it an ideal tool to furthermore refurbish a smartphone for different applications.
While Buildroot provides a default configuration and support for the Pine64 board (based on the same A64 SoC as the PinePhone), peripheral mappings differ, requiring additional adjustments to run on the PinePhone.

Configuring Buildroot introduced additional complexity, as several parameters had to be adapted to match the PinePhone hardware, including kernel options, bootloader integration, and root filesystem generation. 
Although much of the required information was available in the DTS, certain memory mappings and peripheral configurations still had to be identified and specified manually.

Because a functional image had already been produced in previous experiments, it was possible to compare configuration logs between the manual compilation process and the Buildroot workflow.
This comparison facilitated the identification of the specific parameters required by Buildroot, making the overall configuration process more manageable.

Overall, this experiment did not improve the hardware-backed security compared to the previous minimal U-Boot experiment.
It achieves roughly the same result—a fully controllable Linux system on the PinePhone—but demonstrates a more reproducible and flexible workflow that can be reused for other platforms or configurations, as long as critical information (such as the DTS) is publicly available.

\subsection{Attempted TEE Integration}

The final experiment aimed to integrate a Trusted Execution Environment using the open-source OP-TEE implementation into the Buildroot-generated Linux system on the PinePhone.
The goal was to re-enable hardware-backed isolation and provide secure services, restoring some of the security guarantees typically offered by mainstream smartphones.

This step proved significantly more challenging than the previous experiments.
The main difficulties arose from the required low-level integration and configuration of OP-TEE: properly mapping the TEE's secure memory regions and specifying its safe memory zones proved complex, and incorrect mappings caused the Linux kernel to crash.
While OP-TEE itself is well-supported by the kernel, adapting it to the PinePhone's memory layout and boot sequence required intricate configuration that could not be fully resolved.

Despite careful debugging and repeated adjustments, a fully operational OP-TEE system could not be achieved.
The typical breaking point is related to incorrect memory mapping issues (proper memory regions could not be identified), leading to unauthorized access from the kernel to the secure world and, in the end, kernel crashes.

This experiment illustrates that even on an open platform like the PinePhone, achieving a complete TEE integration requires substantial low-level expertise and precise hardware configuration.
While this experiment was a key step to meaningfully enhance hardware-backed security in a repurposed smartphone, it could not be achieved.
The difficulties  encountered did not stem from inherent platform limitations, but from the  high complexity of low-level configuration, including secure memory mapping and safe zone setup, even with complete access to the SoC documentation and the DTS.

\subsection{Summary of Observations}

The sequence of experiments provides a structured view of what can and
cannot be achieved with reasonable work when repurposing a PinePhone.

\begin{figure*}
    \centering
    \includegraphics[width=0.95\linewidth]{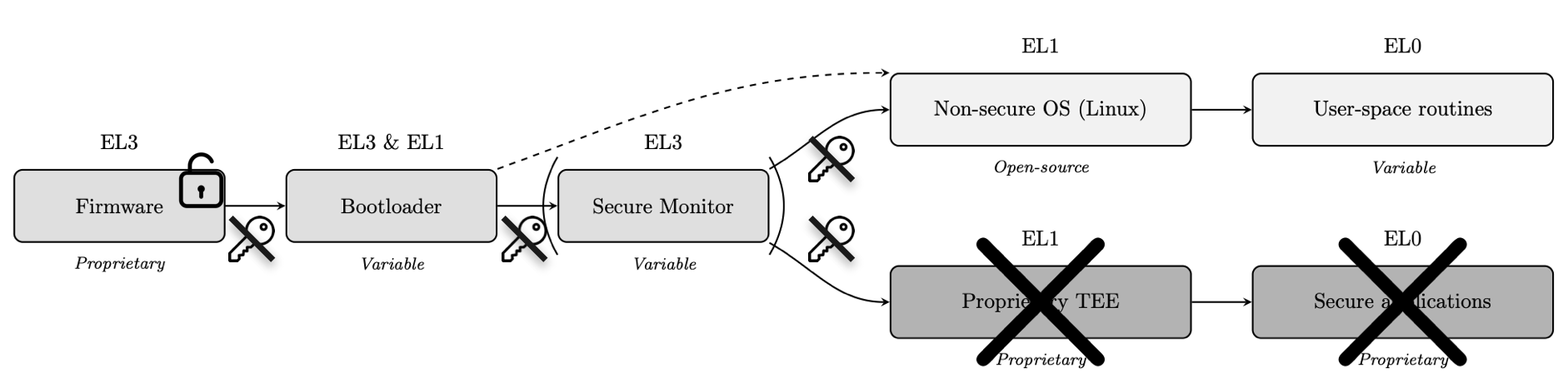}
    \caption{Software Stack achieved on a PinePhone}
    \label{fig:bc_final}
\end{figure*}

Figure~\ref{fig:bc_final} illustrates the resulting boot chain, indicating
which stages were successfully implemented and which remain inaccessible.
The results reveal a clear gap between the ability to deploy a functional
system and the ability to restore hardware-backed security guarantees.

Overall, while repurposing entails full control over the software stack and allows deployment of flexible embedded systems, re-establishing security properties such as verified boot and trusted execution remains a complex and unresolved challenge.
Failing to achieve a fully operational TEE highlights  a fundamental obstacle: even on an open platform like the PinePhone, restoring strong security guarantees in repurposed devices requires substantial expertise and a significant amount of time.
This constitutes a structural barrier to using reconditioned smartphones for applications that demand robust hardware-enforced protections.
Furthermore, the lack of open and detailed documentation about the SoC from the manufacturer is a significant obstacle: the DTS is required for every piece of software replaced, and achieving the same results on a closed-source platform would require a tedious reverse-engineering work beforehand.

%% file: bc_typ.tex
\begin{tikzpicture}[
  font=\small,
    inibox/.style={
    draw,
    thick,
    rounded corners,
    minimum width=3.5cm,
    minimum height=1.2cm,
    align=center,
    fill=gray!25
  },
  rbox/.style={
    draw,
    thick,
    rounded corners,
    minimum width=5cm,
    minimum height=1.2cm,
    align=center,
    fill=gray!10
  },
  secbox/.style={
    draw,
    thick,
    rounded corners,
    minimum width=5cm,
    minimum height=1.2cm,
    align=center,
    fill=gray!60
  },
  labelstyle/.style={
    font=\scriptsize\itshape,
    align=center
  },
  >=stealth
]

% Étapes horizontales
\node[inibox] (fw) {Firmware};
\node[inibox, right=1cm of fw] (bl) {Bootloader : U-boot};
\node[inibox, right=1cm of bl] (sm) {Secure Monitor : TF-A};

% Split en deux colonnes après Secure Monitor, décalées vers la droite
\node[rbox, right=2cm of sm, yshift=1.6cm] (os) {Non-secure OS (Linux)};
\node[secbox, right=2cm of sm, yshift=-1.6cm] (tee) {OP-TEE};

% User-space / Secure applications
\node[rbox, right=1cm of os] (usr) {User-space routines};
\node[secbox, right=1cm of tee,] (secapp) {Secure applications};

% Flèches avec courbes plus naturelles
\draw[->, thick] (fw.east) -- (bl.west);
\draw[->, thick] (bl.east) -- (sm.west);
\draw[->, thick] (sm.north east) to[out=30,in=180] (os.west);
\draw[->, thick] (sm.south east) to[out=-30,in=180] (tee.west);
\draw[->, thick] (os.east) -- (usr.west);
\draw[->, thick] (tee.east) -- (secapp.west);

\node[label, above=2mm of fw.north] {EL3};
\node[label, above=2mm of bl.north] {EL3 \& EL1};
\node[label, above=2mm of sm.north] {EL3};
\node[label, above=2mm of os.north] {EL1};
\node[label, above=2mm of tee.north] {EL1};
\node[label, above=2mm of usr.north] {EL0};
\node[label, above=2mm of secapp.north] {EL0};

\node[labelstyle, below=1mm of fw.south] {Proprietary};
\node[labelstyle, below=1mm of bl.south] {Open-source};
\node[labelstyle, below=1mm of sm.south] {Open-source};
\node[labelstyle, below=1mm of os.south] {Open-source};
\node[labelstyle, below=1mm of tee.south] {Open-source};
\node[labelstyle, below=1mm of usr.south] {Variable};
\node[labelstyle, below=1mm of secapp.south] {Open-source};

\end{tikzpicture}

%% file: discussion.tex
\section{Discussion}
\label{sec:discussion}

The experiments presented in this paper aimed to investigate whether
hardware-backed security guarantees in commodity smartphones can be
effectively recovered after repurposing the device outside its original
provisioning ecosystem. Specifically, we focused on evaluating the boot
chain and the feasibility of integrating a TEE. Our findings show
that when performing a full replacement of the software stack, 
restoring strong hardware-enforced protections is extremely complex
and device-specific. This highlights that, when refurbishing a smartphone
with strict security requirements, users cannot rely solely on a complete
software stack replacement, and alternative approaches are
necessary to achieve meaningful guarantees.

\subsection{Challenges in Recovering Hardware-Backed Security}

Our experiments reveal a significant gap between the PinePhone's hardware
capabilities and the security guarantees achievable after repurposing.
Although the platform supports TrustZone and secure boot, these mechanisms
are not enabled by default once outside of manufacturer provisioning and require careful configuration. Restoring
meaningful hardware-backed security depends entirely on the software stack,
the integrator's expertise, and access to detailed documentation. Even
minor misconfigurations in memory mappings, peripheral assignments, or
bootloader settings can prevent the device from correctly enforcing
isolation or executing a TEE.

Integrating a TEE proved
considerably more challenging than deploying a functional Linux system.
Correctly mapping secure memory regions and defining safe zones requires
low-level expertise, and errors can easily lead to kernel crashes or a
non-functional secure world. Furthermore, this integration is strongly
device-dependent: each smartphone model has different memory layouts, boot
stages, peripheral mappings, and hardware quirks. The procedures developed
for the PinePhone cannot be directly reused on other devices, and adapting
them would require substantial reverse engineering and debugging effort.

These challenges demonstrate that the main obstacle to secure smartphone
reuse is not hardware capability, but the complexity and specificity of
software-hardware integration. Even on open platforms with complete
documentation, fully operational hardware-backed isolation remains difficult
to achieve. For less open or mainstream devices, where bootloader sources
or SoC documentation are proprietary or unavailable, the task becomes
practically infeasible without advanced hardware reverse engineering skills.

In practical terms, this means that while functional reuse of smartphones
is relatively straightforward, achieving strong security guarantees such
as verified boot or isolated execution is a high-difficulty task. For example,
deploying a repurposed PinePhone as a small-scale IoT gateway or development
device is feasible, but using it for applications handling sensitive
cryptographic keys or personal data would be risky without a fully
operational TEE and secure boot chain.

\subsection{Trade-offs and Strategies for Secure Smartphone Reuse}

Secure smartphone reuse involves careful trade-offs between user control,
hardware-backed security, and software support. Fully repurposed devices
offer maximal control over the software stack, but hardware-enforced
protections are effectively disabled. Open platforms such as the PinePhone
suggest that integrating advanced features like TEEs should be easier,
yet our experiments show that achieving functional hardware-backed isolation
remains highly complex and uncertain.

Mainstream devices with extended manufacturer support, such as recent Google Pixel
phones offering seven years of security updates, illustrate a practical path
towards maintaining hardware-backed security over a significant portion of the
device's effective lifetime. While this model restricts user control over the
platform, the combination of controlled boot mechanisms and regular security
updates ensures that devices remain secure well beyond the typical three-year
ownership period observed for most smartphones. This extended support window
provides a realistic opportunity to refurbish devices and redeploy them in
new operational contexts without immediately compromising security. However,
once manufacturer support ends, re-establishing secure boot chains or TEE
functionality becomes highly challenging, often requiring low-level expertise
and device-specific integration. Consequently, while extended support does
not eliminate the need for careful reuse strategies, it significantly
reduces the barrier for refurbishing mainstream smartphones with strong
hardware-backed guarantees.

Based on our observations, three broad strategies for secure smartphone
reuse can be identified:

\begin{itemize}
    \item \textbf{No hardware-backed security:} Devices can be quickly
    repurposed with full control over the platform. Hardware-enforced
    protections are disabled, and users must explicitly accept the absence
    of a root of trust. This strategy is sufficient for low-risk
    applications such as generic Linux-based development boards, hobbyist
    projects, or IoT nodes with non-sensitive data.

    \item \textbf{Full hardware security recovery:} Devices can, in principle,
    be configured to restore secure boot and TEE functionality, providing
    maximal security guarantees. However, this approach is highly device-
    specific, requires deep low-level expertise, and success is not assured.
    Large-scale deployment is impractical due to the substantial time and
    effort required for configuration and debugging. This path may suit
    research labs or security-focused refurbishing projects, but not mass
    reuse.

    \item \textbf{Manufacturer-supported security:} Devices with long-term
    vendor support maintain hardware-backed guarantees without requiring
    extensive modifications. User control is limited, but devices benefit
    from extended secure lifetimes. This strategy combines predictable
    security with minimal technical overhead, making it suitable for
    commercial reuse programs or enterprise deployments where security is
    critical.
\end{itemize}

Overall, these strategies illustrate that secure reuse is not a binary
choice but a spectrum of trade-offs between effort, control, and achievable
security guarantees. The technical and device-specific nature of integrating
hardware-backed security means that, for most repurposing scenarios, users
must carefully balance their need for control against their security
requirements and available expertise. In this light, secure smartphone reuse
should be considered not only from a sustainability perspective but also
as a question of risk management, where the intended application and threat
model dictate the appropriate trade-off between flexibility and security.

\subsection{Implications for future Smartphone Ecosystems and Industry
Practices}

Our findings suggest that the challenges of secure smartphone repurposing
are not only technical, but also reflect structural assumptions in current
smartphone ecosystems. Hardware-backed security mechanisms are designed
around manufacturer-controlled provisioning, updates, and certification,
providing strong guarantees within a single supported lifecycle.

However, these design choices make secure repurposing difficult once a
device leaves its original ecosystem. Re-establishing trust anchors or
configuring TEEs requires low-level documentation and provisioning access
that is typically unavailable outside manufacturer control. As a result,
security mechanisms become tightly coupled to vendor infrastructure.

These constraints reflect economic and organizational incentives in the
smartphone industry, where maintaining long-term support incurs non-trivial
costs and market dynamics favor regular device replacement. Security
architectures are therefore optimized for a controlled lifecycle rather
than multi-cycle reuse.

This creates a tension between sustainability goals and current industry
practice. While right-to-repair initiatives, open-source platforms, and
extended support programs aim to prolong device lifetimes, our results show
that functional longevity alone is insufficient to preserve hardware-backed
security after repurposing.

Finally, our findings highlight opportunities for technical standardization.
Much of the observed complexity stems from platform-specific boot and
security integration mechanisms. Standardized interfaces for secure boot,
TEEs, and ownership transfer could significantly reduce barriers to secure
reuse, even if full hardware uniformity is unrealistic.

At the policy level, extending minimum software support periods appears
more effective for preserving both security and sustainability than
post-hoc repurposing efforts.

In earlier LIMITS literature, M\"uhlberg~\cite{muhlberg_sustaining_2022}
argues that \enquote{ICT systems that sustainably incorporate safety and
security are designed to minimize the risk of hazards for all parties
involved in the life cycle of the system, while maximizing the safe and
secure life span and possibilities for reuse of that system and its
components.} Technical requirements for sustainable safety and security
have been laid out in~\cite{muhlberg_sustaining_2022} and earlier by Paverd
et al.~\cite{paverd_sustainable_2019}, and comprehensively address changing
system requirements, failure and change in third-party systems or the
maintenance environment, software failures and subsystem compromise, and
others.

These requirements strongly link to repurposing scenarios where, as we
illustrate and argue in this paper, devices are taken out of the ecosystem
they were designed for, facing entirely new requirements and operating
environments. As M\"uhlberg and Paverd et al. argue, there are strategies
and design principles -- technical and organizational -- to prepare a device
for these changes. Following the ideas
from~\cite{muhlberg_sustaining_2022,paverd_sustainable_2019} would not only
be to the benefit of reuse and repurposing initiatives but generally enable
stronger and more sustainable notions of security and safety, of
preparedness for incidents and for recovery after incidents, for consumer
equipment and without necessarily increasing system complexity. This again
intersects with the idea of preparedness in \enquote{Resistance
Technologies} (Ben Guirat \& M\"uhlberg,~\cite{guirat_resistance_2025}) and
renunciation and digital de-escalation (Girard et
al.,~\cite{girard_computing_2025}), which both question the compatibility
of current economic models and monetization strategies in the ICT sector
with requirements for foresight and planning in our societies to engage
with an ecological and climate polycrisis and the political impacts of this
polycrisis~\cite{knowles_climate_2025}. This paper contributes to the
discussion around preparedness and sustaining security on repurposed
personal devices by experimentally exploring the opportunities and
highlighting the need for repurposing and reuse scenarios to be considered
in the design phase of new devices to enable secure \enquote{junkyard
computing} (paraphrasing~\cite{switzer_junkyard_2023}
and~\cite{snodgrass_windternet_2024}) with reduced ecological impacts.

%% file: conclusions.tex
\section{Conclusions}
\label{sec:conclusions}

Smartphones are one of the most widely used types of consumer electronics
globally. Following recent trends, 20\,\% of the world population will
purchase a new smartphone in
2026.\footnote{Statista: Smartphones -- Worldwide, \url{https://www.statista.com/outlook/cmo/consumer-electronics/telephony/smartphones/worldwide}}
Given the proliferation, versatility, compute power, and the relatively
short lifetime of these devices, ideas and narratives towards notions of
environmental sustainability and the general reduction of environmental
impact of the smartphone industry have been raised by academics and society
at large. Backed by data from life-cycle assessments, extending the
lifetime of smartphones through repair, refurbishment, reuse and
repurposing is generally considered to be the most effective strategy for
impact reduction. Research in this direction predominantly focuses on
designing devices for repairability as well as on exploring the feasibility
of reuse and repurposing with different approaches and across different
scenarios.

In our research we seek to understand to what extent repurposed smartphones
can be trusted from a security perspective, whether they can be used in
security-critical applications, and whether and how hardware-backed
security features in commodity phones, specifically Trusted Execution
Environments, can be relied upon (or whether such features may even impede
trustworthy reuse scenarios). We address these questions by
envisioning different use cases and by experimenting with an open-hardware
smartphone to then generalize our findings to less transparent consumer
devices. We report on our work towards restoring trust anchors, and on
enabling mechanisms to protect system integrity and secure credential
storage. We conclude that secure reuse is, in general, technically feasible
across many devices but remains very challenging due to variability across
different smartphone models, lack of support from the device vendor, and a
general lack of tooling and documentation. We hope for future generations
of devices to follow, e.g., right-to-repair legislation, and be designed
with repairability and an extended lifespan in mind to ease security
implementations in reuse and repurposing scenarios. Until this is the case,
we invite researchers to follow our experiments and results to explore and
integrate notions of security and dependability in permacomputing, solar
computing, and in other responsible and limits-aware approaches to
computing.

%% file: acks.tex
\section*{Open Science}
All code developed to create the PinePhone flashed images is available as
an artifact under GPLv3 license at
\url{https://gitlab.ulb.be/aroty/pinephone_exps}.
The artifact was under development at the time of submission of this paper
and will likely be completed during the review period.

\section*{Acknowledgements}
We gratefully acknowledge the Brussels-Capital Region – Innoviris for
financial support under grant numbers 2024-RPF-2 MImPG and 2024-RPF-4 SDM,
and the CyberExcellence program of the Walloon Region of Belgium under
grant number 2110186. This work was supported by the Fonds de la Recherche Scientifique – FNRS under Grant No J.0049.23.